\documentclass[aps,prl,twocolumn,groupedaddress,preprintnumbers]{revtex4}
\def\tr{\mathop{\rm tr}\nolimits}
\def\Tr{\mathop{\rm Tr}\nolimits}

\def\mylimit#1{\mathrel{\mathop{\kern0pt\longrightarrow}\limits_{#1}}}

\makeatletter
\@addtoreset{equation}{section}
\makeatother

\newcommand{\VEV}[1]{\left\langle #1 \right\rangle}

\newcommand{\nn}{\nonumber}

\newcommand{\order}[1]{${\cal O}(#1)$}
\newcommand{\GeV}{\mbox{GeV}}

\newcommand{\sub}[1]{$_{\mrm{#1}}$}
\newcommand{\eff}{{\mrm {eff}}}

\newcommand{\bequ}{\begin{equation}}
\newcommand{\eequ}{\end{equation}}
\newcommand{\beqn}{\begin{eqnarray}}
\newcommand{\eeqn}{\end{eqnarray}}
\newcommand{\bctr}{\begin{center}}
\newcommand{\ectr}{\end{center}}
\newcommand{\Ls}{\left(}
\newcommand{\Rs}{\right)}

\newcommand{\Ll}{\left[}
\newcommand{\Rl}{\right]}
\newcommand{\LL}{\left.}
\newcommand{\RR}{\right.}

\newcommand{\mrm}{\rm}

\begin{document}
\preprint{\vbox{\hbox{hep-ph/0209217}
\hbox{KUNS-1804}
\hbox{\today}}}

%Title of letter
\title{Gauge Coupling Unification in GUT with Anomalous $U(1)$ Symmetry
}

\author{Nobuhiro {\sc Maekawa}}
\email[]{maekawa@gauge.scphys.kyoto-u.ac.jp}
\author{Toshifumi {\sc Yamashita}}
\email[]{yamasita@gauge.scphys.kyoto-u.ac.jp}
%\thanks{}
\affiliation{Department of Physics, Kyoto University, Kyoto 606-8502, Japan}

\date{\today}

\begin{abstract}
We show that in the framework of grand unified theory (GUT) 
with anomalous $U(1)_A$ gauge symmetry, the success of the gauge 
coupling unification in the minimal $SU(5)$ GUT is naturally 
explained, even if the mass spectrum of superheavy fields does not 
respect $SU(5)$ symmetry. 
Because the unification scale for most realizations of the 
theory becomes smaller 
than the usual GUT scale, 
it suggests that the present level of experiments is close
to that sufficient to observe 
proton decay via dimension 6 operators, $p\rightarrow e+\pi$.
\end{abstract}

%\keywords{}

\maketitle

% body of letter here - Use proper section commands
\section{Introduction}

As is well known, if a reasonable SUSY breaking scale is assumed, 
the three gauge coupling constants meet 
at the usual GUT scale $\Lambda_G\sim 2\times 10^{16}$ GeV, 
in the minimal supersymmetric standard model (MSSM).
This is a very significant result, and it is sometimes regarded as 
evidence supporting the validity of the existing supersymmetric 
grand unified theory (SUSY-GUT). 
However, generically in the SUSY-GUT scenario, there exist color triplet 
partners of the MSSM Higgs, whose presence results in too rapid
proton decay via dimension 5 operators.
In many GUTs, suppression of this proton decay is incompatible 
with the success of 
gauge coupling unification~\cite{Murayama,Goto,lattice}.
Of course, it is possible to manipulate these theories in such a way
to specifically realize both suppression of proton decay and gauge coupling
unification - for example, with the theories adjusted so that 
proton decay is appropriately suppressed, scales below $\Lambda_G$
can be tuned to realize gauge coupling unification there - but it is 
obviously desirable to construct a theory in which these effects are realized
in a more natural manner. 
However, if we wish for them
to emerge
naturally within any of the existing theories,
there are few possibilities.
Most existing solutions realize the MSSM below $\Lambda_G$
and include suppression or complete prohibition of 
proton decay~\cite{kawamura,hall,kakizaki,maru,babu}.
However, recently another type of 
solution has been proposed~\cite{Unif} in the context
of GUT with anomalous $U(1)$ gauge symmetry~\cite{U(1)}, whose anomaly 
is cancelled by the Green-Schwarz mechanism~\cite{GS}. It is surprising 
that, with this solution,  gauge coupling unification is realized without
fine tuning of the theory,
even though the unified gauge group $G$ $(=SO(10)$ or $E_6)$ has a higher 
rank
than $SU(5)$ (implying that there are several gauge symmetry 
breaking scales), and the mass spectrum of superheavy fields does not 
respect $SU(5)$ symmetry.
This unification is made possible because the mass spectrum of superheavy 
fields and the symmetry
breaking scales are determined by anomalous $U(1)_A$ charges, and
most of the charges cancel and as a result do not appear in
the relations for gauge coupling unification.

In this letter, we give a group theoretical argument that accounts for
the cancellation of the charges within the solution proposed in
Ref. \cite{Unif}. With the understanding provided by this argument,
it becomes 
obvious that any GUTs with anomalous $U(1)_A$ gauge symmetry
can naturally explain the success of gauge coupling unification 
in the minimal
$SU(5)$ GUT
if the following conditions are satisfied:
\begin{enumerate}
\item The unification group $G$ is simple.
\item The vacuum expectation values (VEVs) of GUT gauge singlet operators 
($G$-singlets)
$O_i$ with anomalous $U(1)_A$ charges $o_i$ are given by
\begin{equation}
\VEV{O_i}\sim \left\{ 
\begin{array}{ccl}
  \lambda^{-o_i} & \quad & o_i\leq 0 \\
  0              & \quad & o_i>0
\end{array} \right. . 
\label{VEV}
\end{equation}
\item Below a certain scale, MSSM is 
realized.
\end{enumerate}
Here $\lambda\ll 1$ is the ratio of the cutoff scale $\Lambda$ and 
the VEV of the Froggatt-Nielsen (FN) field $\Theta$, whose 
anomalous $U(1)_A$ charge is
normalized to $-1$~\cite{FN}. 
Throughout this letter, we denote all the superfields and chiral 
operators by uppercase letters and their anomalous $U(1)_A$ 
charges by the corresponding lowercase letters.
In most situations, we use units in which $\Lambda=1$.
In the argument we give, the concept of the ``effective charge" is 
very important, and in fact when it can be defined consistently,
gauge coupling unification is naturally realized. We show that such
a definition can be given when the VEV structure satisfies (\ref{VEV}).
We thus find that gauge coupling unification emerges naturally
in a whole class of GUT models with anomalous $U(1)_A$ symmetry
if their VEV structure satisfies (\ref{VEV}).
In fact, this class includes both our models considered in
Ref. \cite{TGUT} and those proposed
previously in Ref. \cite{Higgs,Matter}.
Moreover, some of the above conditions can be weakened. 
For example, even when the gauge group
is non-simple, gauge coupling unification is realized if the charge 
assignment
respects $SU(5)$ symmetry.

\section{Vacuum determination and mass spectrum of superheavy fields}
In order to examine the gauge coupling unification conditions, we have to
know how the VEVs of Higgs fields and 
the mass spectrum of superheavy fields are determined by the anomalous 
$U(1)_A$ charges.

First, let us recall the vacuum structure of theories with
anomalous $U(1)_A$ gauge symmetry in which generic interactions, namely all
the interactions allowed by the symmetry, are introduced~\cite{TGUT}. 
Generally, in vacua in which 
the FN mechanism~\cite{FN} acts, the VEVs of $G$-singlets $O_i$, 
which are fixed by $F$-flatness conditions, are as given
in Eq.~(\ref{VEV}). 
The VEVs of Higgs fields
can be evaluated by using those relations for the VEVs of the $G$-singlets.
As an example, let us study $SO(10)$ GUT. 
Consider an adjoint Higgs field $A({\bf 45})$ and a pair of spinor 
Higgs fields $C({\bf 16})$ and $\bar C({\bf \overline{16}})$, 
so that $SO(10)$ is broken into the gauge group of the 
standard model (SM) $G_{\rm SM}$.
Then, the scale of the VEV of $A({\bf 45})$
is given approximately by $\VEV{A}\sim \lambda^{-a}$, because the  
$G$-singlet $\tr A^2$ must have a VEV equal to $\lambda^{-2a}$. 
The scale of the VEV of
the complex Higgs $C$ and $\bar C$ behaves as 
$\VEV{\bar CC}\sim \lambda^{-(c+\bar c)}$, and the $D$-flatness condition 
requires
$|\VEV{C}|=|\VEV{\bar C}|\sim \lambda^{-\frac{1}{2}(c+\bar c)}$.
Note that the VEVs of $G$-non-singlet fields can differ from
the expected values for $G$-singlets, that is $\lambda^{-c}$ and 
$\lambda^{-\bar c}$.

Next, we examine how to determine the mass spectrum of superheavy fields.
The mass term of the vector-like fields $X$ and $\bar X$ can be 
written as
\begin{equation}
\lambda^{x+\bar x}\bar XX,
\label{mass}
\end{equation}
where $x+\bar x\geq 0$. 
It is obvious that if $x+\bar x<0$, this mass term is forbidden by the 
symmetry if the vacuum structure satisfies (\ref{VEV}) 
(the SUSY zero mechanism). 
Note that by causing the VEVs to satisfy Eq.~(\ref{VEV}),
the higher-dimensional terms
$
\lambda^{x+\bar x+o_i}\bar XXO_i
$
also induce the masses of $X$ and $\bar X$ that are of the same order as 
that given in Eq.~(\ref{mass}).  This is one of the most important features
of theories with anomalous $U(1)_A$ symmetry.
However, $G$-non-singlet fields ($C$) have generally VEVs that differ from
the expected values for $G$-singlets (i.e. $\lambda^{-c}$).
For this reason, the masses resulting from the VEVs of $G$-non-singlet
fields are generally different from those obtained by simple sum of
the charges, given in Eq. (\ref{mass}).
For example, by introducing the fields $\Psi({\bf 16})$ and $T({\bf 10})$, 
from the interaction 
$
\lambda^{\psi+t+c}\Psi T C,
$
we can obtain the mass of ${\bf \bar 5}_\Psi$ and ${\bf 5}_T$ of
$SU(5)$ as
$
\lambda^{\psi+t+c}\VEV{C}\sim \lambda^{\psi+t+\frac{1}{2}(c-\bar c)},
$
which is generally not equal to the simply expected value,
$\lambda^{\psi+t}$.
In such cases, 
the charges appearing in the mass matrices of superheavy fields can be
replaced by effective charges,
which are defined so that the relations
\begin{equation}
     \VEV{C}\sim \lambda^{-\tilde c}, \VEV{\bar C}
\sim \lambda^{-{\tilde{\bar c}}}
\label{effVEV}
\end{equation}
are satisfied. 
Here, the effective charges $\tilde c $ and $\tilde {\bar c}$ are
related as
$
\tilde c=c+\Delta c=\tilde {\bar c}=\bar c-\Delta c=\frac{1}{2}(c+\bar c),
$
with the discrepancy between these and the original charges given by
$
\Delta c\equiv \frac{1}{2}(\bar c-c).
$
This discrepancy can be understood as the effect of 
an additional $U(1)_V$ in the decomposition
$SO(10)\rightarrow SU(5)\times U(1)_V$.
The VEVs $|\VEV{C}|=|\VEV{\bar C}|$ that break $U(1)_V$
become the source of the new hierarchical structure~\cite{Ramond}.
The unit of this new hierarchy is given by 
$\lambda^{\Delta c}$ $(\equiv\lambda^{\Delta_V})$. (Here,
we normalize the $U(1)_V$ charge of $\VEV{C}$ to 1.)
We can include the effect of the new hierarchical source by
defining the effective charges $\tilde f$ for other fields $F$, 
using $U(1)_V$ charges $v_f$ as
\begin{equation}
\tilde f \equiv f+v_f\Delta_V.
\end{equation}
It is obvious that this definition of the effective charges does not
change the VEV relation in Eq.~(\ref{VEV}), because the $G$-singlets
$O_i$ have vanishing $U(1)_V$ charges.
Note that the effective charges respect $SU(5)$ symmetry,
because $U(1)_V$ respects this symmetry.
The extension of the concept of effective charges to a more general 
situation is straightforward.
If there are several Higgs fields that break $U(1)_V$, the unit
of the new hierarchy can be defined by the Higgs fields with the largest
VEVs. In the case that there are several $U(1)_k$ with the GUT gauge 
group 
$G\rightarrow SU(5)\times \prod_k U(1)_k$,
the unit of the new hierarchy, $\lambda^{\Delta_k}$, can be defined for 
each $U(1)_k$ from the Higgs fields with the largest VEVs that 
break $U(1)_k$. 
Also, by defining the effective charges as 
$\tilde x_i\equiv x_i+\sum_k v_{x_i}^k \Delta_k$ for the superheavy 
fields $X_i$ with $U(1)_k$ charges $v_{x_i}^k$, their masses are 
easily evaluated as
\begin{equation}
\lambda^{\tilde x_i+\tilde x_j},
\end{equation}
unless the mass terms are forbidden by some mechanism, such as the 
SUSY zero mechanism.
Therefore, the determinants of the mass matrices $M_I$ of superheavy 
fields, which appear in the expressions of the gauge coupling flows, 
are written 
$
\det M_I = \lambda^{\Sigma_i \tilde x_i^I},
$
where $I$ is the index for the SM irreducible representations.
Note that $\det M$ can be calculated using the simple sum of 
the effective charges of the massive fields.
The ratio of the determinants for each pair of SM multiplets
$I$ and $I'$ (contained in each $SU(5)$ multiplet),
$\frac{\det M_I}{\det M_{I^\prime}}$,
appears in the relations for gauge coupling unification.
Because the effective charges respect $SU(5)$ symmetry, 
the contributions of the $SU(5)$ multiplet whose $I$ and 
$I^\prime$ components are both massive cancel.
Hence, only the effective charges of massive modes 
whose $SU(5)$ partners are massless contribute.
This can be reinterpreted as meaning that only the effective charges of the 
massless modes appear in the ratios, that is,
\bequ
  \frac{\det M_I}{\det M_{I^\prime}}
 = \frac{1/\lambda^{\Sigma_i \tilde x_i^I}}
        {1/\lambda^{\Sigma_i \tilde x_i^{I^\prime}}}, 
\label{ratio}
\eequ
where $i$ runs over the massless modes.

\section{Gauge Coupling Unification}
Now, we carry out an analysis based on the renormalization group
equations (RGEs) up to one loop. 
Here, we consider the most general situation, in which the GUT symmetry $G$ 
is successively broken into $G$\sub{SM} as
\bequ
  G(\equiv H_0)
   \mylimit{\Lambda_1}H_1
   \mylimit{\Lambda_2}\dots
   \mylimit{\Lambda_N}G_{\rm{SM}}
                                   (\equiv H_N).
\eequ
First, the conditions for the gauge coupling unification are given by
$
\alpha_3(\Lambda)=\alpha_2(\Lambda)=
\frac{5}{3}\alpha_Y(\Lambda)\equiv\alpha_1(\Lambda),
$
and the gauge couplings at the cutoff scale $\Lambda$ are given by
\beqn
\alpha_a^{-1}(\Lambda)&=&\alpha_a^{-1}(M_{SB})
  +\frac{1}{2\pi}\Ls b_a\ln\Ls\frac{M_{SB}}{\Lambda}\Rs\RR\nn\\
   &&\hspace{-0.8cm}\LL +\sum_i \Delta b_{ai}\ln\Ls\frac{m_i}{\Lambda}\Rs
+\sum_n \Delta_{an}\ln\Ls\frac{\Lambda_n}{\Lambda}\Rs
                 \Rs\,,
\eeqn
where $a=1,2,3$, $M_{SB}$ is the SUSY breaking scale, 
$(b_1,b_2,b_3)=(33/5,1,-3)$ are the 
renormalization group coefficients of MSSM,
$\Delta b_{ai}$ are the corrections to the coefficients 
caused by the massive fields with masses $m_i$,
 and the last term 
is the correction due to the enhancement of the gauge symmetry 
above each symmetry breaking scale $\Lambda_n$:
\bequ
  \Delta_{an}=-3T_a\Ll H_{n-1}/H_n\Rl 
              +T_a\Ll\mbox{NG}_n\Rl=-2T_a\Ll\mbox{NG}_n\Rl.
\label{Delta}
\eequ
Here, NG$_n$ denotes the NG modes that are absorbed through the Higgs 
mechanism at the scale 
$\Lambda_n$, and $T_a$ are the Dynkin indices, defined as 
$
  \Tr(T_AT_B)=T[R]\delta_{AB}, 
$
where $T_A$ are the generators in the $R$ representation.
The second equality in Eq.~(\ref{Delta}) is derived from the
relation $T_a[H_{n-1}/H_n]=T_a[\mbox{NG}_n]$.

Then, using the fact that in MSSM the three gauge couplings meet at 
the scale $\Lambda_G\sim2\times10^{16}\GeV$, the relations expressing 
unification, $\alpha_a(\Lambda)=\alpha_b(\Lambda)$, become
\beqn
  (b_a-b_b)\ln(\Lambda_G)
  &+&\sum_I(\Delta b_{aI}-\Delta b_{bI})\ln\Ls\det M_I\Rs
  \nn\\
  &+&\sum_n(\Delta_{an}-\Delta_{bn})\ln\Ls\Lambda_n\Rs=0,
\label{Cond}
\eeqn
where $I$ runs over the SM irreducible representations.
Because the sum of $\Delta b_{aI}$ over an $SU(5)$ 
multiplet is independent of $a$, the second term in Eq. (\ref{Cond})
can be written in terms of 
the ratios of the determinants of the mass matrices in (\ref{ratio}), 
and therefore in terms of the contributions from the massless modes, 
as mentioned above.
In terms of the ``effective mass'' of massless modes, which is defined 
as $m_\eff\equiv\lambda^{\tilde x+ \tilde y}$, even if 
$\tilde x+ \tilde y<0$, the second term in (\ref{Cond}) can be written
\bequ
  \sum_{i=\mbox{\scriptsize{massless}}}
  (T_a\Ll i\Rl -T_b\Ll
          i
      \Rl)
  \ln\Ls 1/m_\eff^{i}\Rs.
\nn
\eequ
These massless modes consist of two types, physical massless modes, 
such as the MSSM doublet Higgs ($H_u$ and $H_d$), and unphysical NG modes.
From (\ref{Delta}), we can see that the contribution of the latter 
type is cancelled by that of the last term in Eq.~(\ref{Cond})
if the conditions  
\bequ
  m_\eff^{\mbox{\scriptsize{NG}}_n}\sim\Lambda_n^{-2}
\label{condi}
\eequ
hold.
These conditions are satisfied when the 
vacuum structure satisfies (\ref{VEV}),
because $m_\eff^{\mbox{\scriptsize{NG}}_n}$ is the coefficient of 
the bilinear term of the $n$-th NG modes, 
$\Phi$ and $\bar\Phi\ (\tilde {\bar\phi}=\tilde \phi)$, 
and therefore 
$m_\eff^{\mbox{\scriptsize{NG}}_n}\sim\lambda^{2\tilde \phi}$, 
and from (\ref{effVEV}), $\Lambda_n\sim\lambda^{-\tilde \phi}$.

When (\ref{condi}) holds, only the physical massless 
modes contribute to the conditions for the gauge coupling 
unification, and they are independent of the details of the Higgs sector, 
such as the field content and the symmetry breaking pattern.
In particular, if all the fields other than those in MSSM become 
superheavy, only the MSSM doublet Higgs fields $H$ contribute, and
we have 
\bequ
  (b_a-b_b)\ln(\Lambda_G)
  +(\Delta b_{aH}-\Delta b_{bH})\ln\Ls {1/m_\eff^H}\Rs =0,
\eequ
for all combinations $(a,b)$. These relations lead to 
$\ln(\Lambda_G)=\ln(m_\eff^H) =0$, and thus 
\bequ
 \quad\Lambda\sim \Lambda_G\,,\,\,\tilde h_u+\tilde h_d\sim 0.
\label{unification}
\eequ
The first relation here simply defines the scale of the theory:
The cutoff scale $\Lambda$ is taken as the usual GUT scale, 
$\Lambda_G$.
This is also the case in the minimal $SU(5)$ GUT, where
the scale at which $SU(5)$ is 
broken is also taken as $\Lambda_G$. 
The second relation in (\ref{unification}) corresponds to that for the 
colored Higgs mass in the minimal $SU(5)$ GUT, because the effective 
colored Higgs mass is obtained as 
$m^{H^c}_\eff\sim \lambda^{\tilde h_u+\tilde h_d}$. 
Therefore, we have no tuning parameters for the gauge coupling 
unification other than those in the minimal $SU(5)$ GUT.
Note that (when $\tilde h_u+\tilde h_d= 0$) 
if we calculate gauge couplings at a low energy scale in
the GUT scenario with any cutoff (for example, the Planck scale)
and use them as the initial values, the three running gauge
couplings calculated in MSSM meet at the cutoff scale.
In this way, we can naturally explain 
the gauge coupling unification in the minimal $SU(5)$ GUT.

Note that the relation $\tilde h_u+\tilde h_d\sim 0$ does not 
imply $\tilde h_u+\tilde h_d=0$, 
 because there is an ambiguity involving
\order1 coefficients, and we have used only one loop RGEs. 
Using the VEV $\VEV{A}$, which breaks $SU(5)$ symmetry, the 
contribution to the mass of $X$ and $\bar X$ from higher-dimensional
interactions, $\lambda^{x+\bar x+na}\bar XA^n X$, is of the same order
as that from
the mass term $\lambda^{x+\bar x}\bar XX$, because 
$\VEV{A}\sim \lambda^{-a}$.
Therefore the \order1 coefficients do not respect $SU(5)$ symmetry. 
This contrasts with the usual situation, in which the contribution from
higher-dimensional interactions is suppressed.
The fact that the \order1 coefficients do not respect $SU(5)$ symmetry 
allows a non-zero value of $\tilde h_u+\tilde h_d$.
 This is important, because 
it is necessary for $\tilde h_u+\tilde h_d$ to be negative 
in order 
to suppress proton decay via dimension 5 operators, 
the order of whose coefficients is determined by their 
effective charges and is independent of their origin 
({\it e.g.} colored Higgs exchange or new physics 
around the cutoff scale).
The suppression 
requires the effective colored 
Higgs mass 
$m_{\rm eff}^{H^c} \sim \lambda^{\tilde h_u+\tilde h_d}\Lambda$
$>$ \order{10^{18}\GeV}, 
and therefore $\tilde h_u+\tilde h_d\leq -3$ is needed~\cite{Unif}.
Note that $m_{\rm eff}^{H^c}>\Lambda$, but the physical masses of the 
colored Higgs are smaller than $\Lambda$. 

\section{Discussion and Summary}
The argument given here is strongly dependent on the vacuum structure 
(\ref{VEV}), which is naturally realized in the GUT scenario 
with anomalous
\footnote{
We can use non-anomalous $U(1)$ symmetry instead of anomalous $U(1)$
symmetry if 3 conditions in the Introduction are satisfied. 
However, since we don't know such models, we adopt anomalous $U(1)$ 
symmetry in this letter.
}
$U(1)_A$ gauge symmetry~\cite{TGUT}.
Our results can also be applied to previously studied GUT scenarios
 with anomalous
$U(1)_A$ symmetry~\cite{Higgs,Matter}, 
distinct from that considered here,
if all the VEVs of the $G$-singlets satisfies the VEV relations 
(\ref{VEV}),
even in the case that there are some flat directions.

We have shown that in the more general framework of GUT with 
anomalous $U(1)_A$ gauge symmetry than in Ref.~\cite{TGUT}, 
the success of gauge coupling unification 
in the minimal $SU(5)$ GUT is naturally explained.
Usually, if we adopt a simple group whose
rank is higher than that of the standard gauge group (for example, 
$SO(10)$,
$E_6$, $SU(6)$, etc.), gauge coupling unification can always 
be realized by  tuning the additional degrees of freedom related with 
the several scales of Higgs VEVs.
However, we have shown that in the framework,
all the charges of the Higgs fields, except that of the MSSM doublet 
Higgs, are cancelled in the relations for gauge coupling unification 
(\ref{unification}),
and therefore we have no tuning parameters for the gauge 
coupling unification other than those in the minimal $SU(5)$ GUT. 
We have elucidated the conditions for this cancellation with a
group theoretical proof.

In the GUT scenario with anomalous $U(1)_A$ symmetry, the unification 
scale is given by $\Lambda_A\sim \lambda^{-a}\Lambda$, 
where $a\leq 0$ is the charge
of the Higgs field whose VEV breaks the usual $SU(5)$ gauge symmetry. 
Because the realization of gauge coupling unification requires 
the cutoff scale to be taken as the usual GUT scale, 
$\Lambda_G\sim2\times10^{16}\GeV$, 
the negative charge $a$ leads to a unification scale smaller
than $\Lambda_G$. Therefore, proton decay via dimension 6 operators 
can be enhanced, although that via the dimension 5 is suppressed
when $\tilde h_u+\tilde h_d\leq -3$.
If we choose $a=-1$ and $\lambda\sim 0.22$ as typical values
\footnote{
Note that the slowest proton decay via dimension 6 operators
is obtained when $a=0$, and
the value must be the same as that in the usual GUT scenario. 
However, when $a=0$, generally terms of the form
$\int d^2\theta A^n W_\alpha W^\alpha$ are allowed,
where $W_\alpha$ is a SUSY field strength.
This makes it impossible to realize natural gauge coupling unification. 
The most
natural way to forbid these terms is to choose $a$ to be negative,
which leads to a shorter proton lifetime.},
the proton lifetime can be roughly estimated, 
using a formula in Ref.~\cite{Murayama} and a recent result provided
by a lattice calculation for the hadron matrix element parameter
$\alpha$~\cite{lattice},
as
{\small{
\begin{equation}
\tau_p(p\rightarrow e\pi^0)\sim 5\times 10^{33}
\left(\frac{\Lambda_A}{5\times 10^{15}\ {\rm GeV}}\right)^4
\left(\frac{0.015({\rm GeV})^3}{\alpha}\right)^2  {\rm yrs}.
\nn
\end{equation}
}}
This value
is near the present experimental limit~\cite{SKproton}.
Thus, our study in this letter gives a strong motivation to serch 
the proton decay $p\rightarrow e+\pi$ in future experiments.

  N.M. is supported in part by Grants-in-Aid for Scientific 
Research from the Ministry of Education, Culture, Sports, Science 
and Technology of Japan.

\end{document}